\documentclass{ws-mpla}

\begin{document}

\markboth{Hakimov, Turimov,  Abdujabbarov, Ahmedov} {Quantum
Interference Effects in Ho\v{r}ava-Lifshitz Gravity}

\catchline{}{}{}{}{}

\title{Quantum Interference
Effects in Ho\v{r}ava-Lifshitz Gravity}

\author{\footnotesize ABDULLO HAKIMOV}

\address{Ulugh Beg Astronomical Institute, Astronomicheskaya 33,
    Tashkent 100052, Uzbekistan \\
   Institute of Nuclear Physics, Ulughbek, Tashkent 100214, Uzbekistan \\
abdullo@astrin.uz}

\author{BOBUR TURIMOV}

\address{National University of Uzbekistan, Tashkent 100095, Uzbekistan
}

\author{AHMADJON ABDUJABBAROV}

\address{Ulugh Beg Astronomical Institute, Astronomicheskaya 33,
    Tashkent 100052, Uzbekistan \\
   Institute of Nuclear Physics, Ulughbek, Tashkent 100214, Uzbekistan
   \\
   Inter-University Centre for Astronomy \& Astrophysics, Post Bag 4 Pune 411 007, India\\
ahmadjon@astrin.uz}

\author{BOBOMURAT AHMEDOV}

\address{Ulugh Beg Astronomical Institute, Astronomicheskaya 33,
    Tashkent 100052, Uzbekistan \\
   Institute of Nuclear Physics, Ulughbek, Tashkent 100214, Uzbekistan
   \\
   Inter-University Centre for Astronomy \& Astrophysics, Post Bag 4 Pune 411 007, India\\
ahmedov@astrin.uz}

 \maketitle

\pub{Received (Day Month Year)}{Revised (Day Month Year)}

\begin{abstract}
The relativistic quantum interference effects in the spacetime of
slowly rotating object in the Ho\v{r}ava-Lifshitz gravity as the
Sagnac effect and phase shift  of interfering particle in neutron
interferometer are derived. We consider the extension of
Kehagias-Sfetsos (KS) solution~\cite{ks09} in the
Ho\v{r}ava-Lifshitz gravity for the slowly rotating gravitating
object. Using the covariant Klein-Gordon equation in the
nonrelativistic approximation, it is shown that the phase shift in
the interference of particles includes the gravitational potential
term with the KS parameter $\omega$. It is found that in the case
of the Sagnac effect, the influence of the KS parameter $\omega$
is becoming important due to the fact that the angular velocity of
the locally non rotating observer is increased
 in Ho\v{r}ava gravity.
From the results of the recent experiments~\cite{holgeretal} we
have obtained lower limit  for the coupling KS constant as $\omega
\simeq 1.25 \cdot 10^{-25} \rm{cm}^{2}$. Finally, as an example,
we apply the obtained results to the calculation of the UCN
(ultra-cold neutrons) energy level modification in the
gravitational field of slowly rotating gravitating object in the
Ho\v{r}ava-Lifshitz gravity.

\end{abstract}

\ccode{PACS Nos.: 04.50.-h, 04.40.Dg, 97.60.Gb.}

\section{Introduction}

One of the biggest difficulties in attempts toward the theory of
quantum gravity is the fact that general relativity is
non-renormalizable. This would imply loss of theoretical control
and predictability at high energies. In January 2009, Petr
Ho\v{r}ava proposed a new theory of quantum gravity with dynamical
critical exponent equal to $z = 3$ in the UV (Ultra-Violet) in
order to evade this difficulty by invoking a Lifshitz-type
anisotropic scaling at high energy. This theory, often called
Ho\v{r}ava-Lifshitz gravity, is power counting renormalizable and
is expected to be renormalizable and unitary~\cite{h1,h2}.

Having a new candidate theory for quantum gravity, it is important
to investigate its astrophysical and cosmological implications.
Thus the Ho\v{r}ava theory has received a great deal of attention
and since its formulation various properties and characteristics
have been extensively analyzed, ranging from formal
developments~\cite{Visser,Sotiriou,H_prl,Cai,Orlando,SotiriouJHEP,Calcagni,Germani},
cosmology~\cite{Takashi,CalcagniJHEP,Kalyana,Wang,loboooo,mukohayama},
dark energy~\cite{Saridakis,parkkkk}, dark
matter~\cite{Mukohyama}, and spherically symmetric~\cite{loboetal}
or axial symmetric
solutions~\cite{ghodsi,CaiCao,Konoplya22,SChen,CastilloLar,ChenYang}.

In the paper Ref.~\refcite{lobo1} the possibility of
observationally testing Ho\v{r}ava gravity at the scale of the
Solar System, by considering the classical tests of general
relativity (perihelion precession of the planet Mercury,
deflection of light by the Sun and the radar echo delay) for the
Kehagias-Sfetsos asymptotically flat black hole
solution~\cite{ks09} of Horava-Lifshitz gravity has been
considered. The stability of the Einstein static universe by
considering linear homogeneous perturbations in the context of an
Infra-Red (IR) modification of Ho\v{r}ava gravity has been studied
in the paper~\cite{lobo2}. In the paper Ref.~\refcite{konoplya11}
author considered potentially observable properties of black holes
in the deformed Ho\v{r}ava-Lifshitz gravity with Minkowski vacuum:
the gravitational lensing and quasinormal modes.

The role of the tidal charge in the orbital resonance model of
quasiperiodic oscillations in black hole systems~\cite{stuklik}
and in neutron star binary systems~\cite{stuklik2} have been
studied intensively.
{ The motion of test particles around black hole immersed in
uniform magnetic field in Ho\v{r}ava gravity and influence of
$\omega$ parameter on radii of innermost stable circular orbit
have been studied in papers Ref.~\refcite{aha hl,gwak}.}

The experiment to test the effect of the gravitational field of
the Earth on the phase shift in a neutron interferometer were
first proposed by Overhauser and Colella~\cite{overhauser}. Then
this experiment was successfully performed by Collela, Overhauser
and Werner~\cite{colella}. After that, there were found other
effects, related with the phase shift of interfering particles.
Among them the effect due to the rotation of the
Earth~\cite{page,werner}, which is the quantum mechanical analog
of the Sagnac effect, and the Lense-Thirring
effect~\cite{mashhoon} which is a general relativistic effect due
to the dragging of the reference frames. So we do not consider the
neutron spin in this paper.

In the paper Ref.~\refcite{kuroiwa} a unified way of study of the
effects of phase shift in neutron interferometer was proposed.
Here we extend this formalism to the case of slowly rotating
stationary gravitational fields in the framework of
Ho\v{r}ava-Lifshitz gravity  in order to derive such phase shift
due to either existence or nonexistence of the KS parameter
$\omega$.

The Sagnac effect is well known and thoroughly studied in the
literature, see e.g. paper~Ref.~\refcite{G.rizzi}. It presents the
fact that between light or matter beam counter-propagating along a
closed path in a rotating interferometer a fringe shift $\Delta
\phi$ arises. This phase shift can be interpreted as a time delay
$\Delta T$ between two beams, as it can be seen below, does not
include the mass or energy of particles. That is why we may
consider the Sagnac effect as the "universal" effect of the
geometry of space-time, independent of the physical nature of the
interfering beams. Here we extend the recent results obtained in
the papers~ Ref.~\refcite{rizzi,ruggiero} where it has been shown
a way of calculation of this effect in analogy with the
Aharonov-Bohm effect, to the case of slowly rotating compact
object in Ho\v{r}ava-Lifshitz gravity.

 In this
paper we study quantum interference effects in particular the
Sagnac effect and phase shift effect in a neutron interferometer
in the Ho\v{r}ava model which is organized as follows. In section
\ref{2ndsec}, we start from the covariant Klein-Gordon equation in
the Ho\v{r}ava model and consider terms of the phase difference of
the wave function. Recently Granit experiment~\cite{nesvizhevsky}
verified the quantization of the energy level of ultra-cold
neutrons (UCN) in the Earth's gravity field and new, more precise
experiments are planned to be performed. Experiments with UCN have
high accuracy and that is the reason to look for verification of
the gravitational effects in such experiments. In this section as
an example we investigate modification of UCN energy levels caused
by the existence of KS (Kehagias and Sfetsos) parameter $\omega$ .
In section \ref{3rdsec} we consider interference in Mach-Zender
interferometer and in Section \ref{4thsec} we study the Sagnac
effect in the background  spacetime of slowly rotating object in
Ho\v{r}ava gravity.

Throughout, we use space-like signature $(-, +, +, +)$,
geometrical units system (However, for those expressions with an
astrophysical application we have written the speed of light
explicitly.). Greek indices are taken to run from 0 to 3 and Latin
indices from 1 to 3; covariant derivatives are denoted with a
semi-colon and partial derivatives with a comma.

\section{The Phase shift \label{2ndsec}}

The four-dimensional metric of the spherical-symmetric spacetime
written in the ADM
formalism~\cite{loboetal,lobo1,lobo2,konoplya11} has the following
form:
\begin{equation}\label{metri}
ds^{2}=-N^{2}c^{2}dt^{2}+g_{ij}(dx^{i}+N^{i}dt)(dx^{j}+N^{j}dt)\ ,
\end{equation}
where $N$, $N^i$ are the metric functions to be defined.

The IR (Infrared) - modified Horava action is given by (see for
more details Ref.~\refcite{h1,h2,loboetal,lobo1,lobo2,konoplya11})
\begin{eqnarray}
&& S = \int dtdx^3 \sqrt{-g} N \bigg[\frac{2}{\kappa
^2}(K_{ij}K^{ij}-\lambda_g
K^2)-\frac{\kappa^2}{2\nu_g^4}C_{ij}C^{ij}+\frac{
\kappa^2\mu}{2\nu_g^2}\epsilon^{ijk}R_{il}\nabla_j
R^l_{\ k} \nonumber\\
&& -\frac{\kappa^2
\mu^2}{8}R_{ij}R^{ij}+\frac{\kappa^2\mu^2}{8(3\lambda_g-1)}\left(\frac{4\lambda_g-
1}{4}R^2-\Lambda_WR+3\Lambda^2\right)+\frac{
\kappa^2\mu^2\omega}{8(3\lambda_g-1)}R\bigg]  \ ,
\end{eqnarray}
where $\kappa, \lambda_g, \nu_g, \mu, \omega$ and $\Lambda_W$ are
constant parameters, the Cotton tensor is defined as
\begin{equation}
C^{ij}=\epsilon^{ikl}\nabla_k (R^{j}_{\ l}-\frac{1}{4} R\delta^j_l
)\ ,
\end{equation}
$R_{ijkl}$ is the three-dimensional curvature tensor, and the
extrinsic curvature $K_{ij}$ is defined as
\begin{equation}
K_{ij}=\frac{1}{2N}(\dot{g}_{ij}-\nabla_i N_j -\nabla_j N_i)\ ,
\end{equation}
where dot denotes a derivative with respect to $t$.

Imposing the case $\lambda_g=1$, which reduces to the action in IR
limit, one can obtain the Kehagias and Sfetsos (KS) asymptotically
flat solution~\cite{ks09} for the metric outside the gravitating
spherical symmetric object in Horava gravity:
\begin{eqnarray}
&& ds^{2}=-N^{2}c^{2}dt^{2}+N^{-2}d
r^2+r^{2}d\theta^{2}+r^{2}\sin^{2}\theta d\varphi^{2},\\
&& N^2=1+\omega r^2 -\sqrt{r(\omega^2 r^3+4\omega M)}, \nonumber
\end{eqnarray}
where $M$ is the total mass, $\omega$ is the KS parameter and the
constant $\Lambda_W=0$ is chosen.

Up to the second derivative terms in the action, one can easily
find the known topological rotating solutions given in
Ref.~\refcite{ghodsi,four}. This metric in the slow rotation limit
has the form:
\begin{equation}\label{metrik}
ds^{2}=-N^{2}c^{2}dt^{2}+N^{-2}d
r^2+r^{2}d\theta^{2}+r^{2}\sin^{2}\theta
d\varphi^{2}-2(1-N^2)ac\sin^{2}\theta dtd \varphi\ ,
\end{equation}
here $a$ is the specific angular momentum of the gravitating
object.

Using the Klein-Gordon equation
\begin{eqnarray}
&& \nabla^{\mu}\nabla_{\mu}\Phi-(mc/ \hbar)^{2}\Phi=0 ,
\end{eqnarray}
for particles with mass $m$ one cane define the wave function
$\Phi$ of interfering particles as~\cite{kuroiwa}
\begin{eqnarray}
&& \Phi=\Psi exp\left(-i\frac{mc^2}{\hbar}t\right) \ ,
\end{eqnarray}
where $\Psi$ is the nonrelativistic wave function.

In the present situation, both parameters $GM/rc^{2}$ and $a/r$
are sufficiently small and their higher order terms can be
neglected. Therefore, to the first order in $M$, $\omega$ and
neglecting the terms of $O((\upsilon/c)^{2})$, the Klein-Gordon
equation in Horava-Lifshitz gravity becomes
\begin{eqnarray}
\label{Schr}&& i\hbar\frac{\partial \Psi}{\partial
t}=-\frac{\hbar^2}{2m}\left[\frac{1}{r^2}\frac{\partial}{\partial
r}\left(r^2\frac{\partial}{\partial
r}\right)-\frac{L^2}{r^2\hbar^2}\right] \Psi+\frac{m c^2\omega
r^2}{2}\left(1-\sqrt{1+\frac{4GM}{ c^2\omega
r^3}}\right)\Psi \nonumber\\
&& \hspace{1.4 cm}-a c\omega \left(1-\sqrt{1+\frac{4GM}{ c^2\omega
r^3}}\right)L_z\Psi\ ,
\end{eqnarray}
where we have used the following notations:
\begin{eqnarray}
&& L^{2}=-\hbar^{2}\left[\frac{1}{\sin\theta}\frac{\partial}
{\partial\theta}\left(\sin\theta\frac{\partial}{\partial\theta}\right)
+\frac{1}{\sin^{2}\theta}\frac{\partial^{2}}{\partial\varphi^{2}}
\right] \ ,
\\
&& L_z=-i\hbar\frac{\partial}{\partial\varphi}\ ,
\end{eqnarray}
which correspond to the square of the total orbital angular
momentum and $z$ component of the  orbital angular momentum
operators  of the particle with respect to the center of the
Earth, respectively.

After the coordinate transformation
$\varphi\rightarrow\varphi+\Omega t$, where
$\Omega=\Omega_{\bigoplus}$ is the angular velocity of the Earth,
we obtain the Schr\"{o}dinger equation for an observer fixed on
the Earth in the following form:
\begin{eqnarray}
\label{Schr} &&i\hbar\Psi_t= H_0\Psi+H_1\Psi+H_2\Psi+H_{3}\Psi \ ,
\end{eqnarray}
where
\begin{eqnarray}
&& H_0=-\frac{\hbar^2}{2m}\frac{1}{r^2}\frac{\partial}{\partial
r}\left(r^2\frac{\partial}{\partial r}\right)+\frac{L^2}{2mr^2}\
,\quad H_1=\frac{m c^2\omega
r^2}{2}\left(1-\sqrt{1+\frac{4GM}{c^2\omega r^3}}\right)\ ,
\quad\nonumber\\
&& H_2=-\Omega L_z \ , \qquad H_3=-a\omega
c\left(1-\sqrt{1+\frac{4GM}{c^2\omega r^3}}\right)L_z .
\label{hamiltonian}
\end{eqnarray}
 $H_{0}$ is the Hamiltonian for a freely propagating particle,
$H_{1}$ is the Horava-Lifshitz gravitational potential energy,
$H_{2}$ is concerned to the rotation, $H_{3}$ is related to the
effect of dragging of the inertial frames. The phase shift terms
due to $H_1, H_2$ and $H_3$ are
\begin{eqnarray}
&&\beta_{Sag}\simeq\frac{2m \bf{\Omega} \cdot\textbf{S}}{\hbar}\ ,
\\
&& \beta_{Hdrag}\simeq\frac{2Gm }{\hbar
c^{2}R^{3}}\textbf{J}\left[\textbf{S}-3\left(1-\frac{2GM}{c^{2}\omega
R^{3}}\right)\left(\frac{\textbf{R}}{R}\cdot\textbf{S}\right)
\frac{\textbf{R}}{R}\right]\ ,
\end{eqnarray}
where ${\bf R}$ represents the position  vector of the instrument
from the center of the Earth, $\mathbf{S}=S\mathbf{n}$,  $S$ is
the area of the interferometer, and $\mathbf{n}$ is the unit
normal vector. If we assume that the Earth is a sphere of radius
$R$ with uniform density then
\begin{equation}
{\bf J}=\frac{2}{5}M R^2{\bf \Omega} \ ,
\end{equation}
and
\begin{equation}
\beta_{\omega}=\frac{\omega R^2}{5}\left(1-\sqrt{1+\frac{4Gm}{
c^2\omega R^{3}}}\right)\beta_{Sag} \ , \quad
\beta_{\omega}=\frac{\omega
R^2}{5}\left(1-\frac{1-\frac{2GM}{c^2\omega
R^{3}}}{\sqrt{1+\frac{4GM}{c^2\omega R^{3}}}}\right)\beta_{Sag}\ ,
\end{equation}
if {\bf R} is perpendicular and parallel to $\bf S$, respectively.
Here $g=GM/R^2$ is the free fall acceleration of Earth.

\begin{figure}[ph]
\centerline{\psfig{file=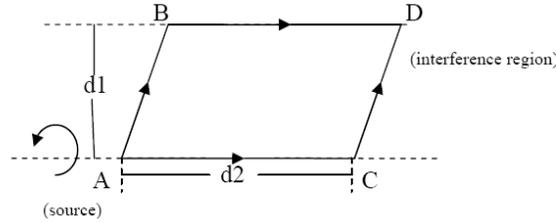,width=3.0in}} \vspace*{8pt}
\caption{Schematic illustration of alternate paths separated in
the vertical direction in a neutron
interferometer.\protect\label{fig1}}
\end{figure}

Now one can easily  calculate the phase shift due to the
gravitational potential. For the purpose of the present
discussion, the quasi-classical approximation is valid and the
phase shift
\begin{eqnarray}
 \beta_{Hgrav}=\beta_{ABD}-\beta_{ACD} \simeq-\frac{1}{\hbar}\int H_{1}dt=\frac{m^2 c^2 S
\lambda\omega R}{2\pi\hbar^2}\left(1-\frac{1+\frac{g}{c^2\omega
R}}{\sqrt{1+\frac{4g}{c^2\omega R}}} \right)\sin\varphi \ , \nonumber\\
\end{eqnarray}
is given by the integration along a classical trajectory.
 Here $ S=d_1d_2$ is the area of interferometer, $\lambda$ is de Broglie
wavelength (see the Fig. \ref{fig1}).

Recently published paper Ref.~\refcite{holgeretal} describes the
precise measurement of the gravitational redshift by the
interference of matter waves in the gravitational field of the
Earth. Comparing their experimental results with our theoretical
predictions one can easily obtain the lower limit on the value of
KS parametr $\omega \simeq 2.4\times10^{-27} \rm{cm}^{-2} \ $.

Astrophysically it is interesting to apply the obtained result for
the Hamiltonian of the particle, moving around rotating
gravitating object in Ho\v{r}ava gravity, to the calculation of
energy level of ultra-cold neutrons (UCN) (as it was done for
slowly rotating space-times in the papers
Ref.~\refcite{arminjon,vika}). The effect of the angular momentum
perturbation of the Hamiltonian $H_2=\Omega L_z$ on the energy
levels of UCN was studied in~\cite{nesvizhevsky} and subsequent
papers. Our purpose is to generalize this correction to the case
of the gravitating object (the Earth in particular case) in
Ho\v{r}ava model. Denote as $\psi$ the unperturbed
non-relativistic stationary state of the 2- spinor (describing
UCN) in the field of the rotating gravitating object in Ho\v{r}ava
gravity. Then we have
\begin{equation}\label{Hamil}
H_{3}\psi =i\hbar a c\omega\left(1-\sqrt{1+\frac{4GM}{\omega
c^2r^3}}\right)\frac{\partial\psi}{\partial\varphi}=i\hbar a
c\omega r\sin\theta\left(1-\sqrt{1+\frac{4GM}{\omega
c^2r^3}}\right)\bf{\nabla}\psi\cdot \bf{e}_{\varphi} \ ,
\end{equation}
here
\begin{equation}
{\bf{\nabla}}\psi=\frac{\partial\psi}{\partial
r}{\bf{e}}_r+\frac{1}{r}\frac{\partial\psi}{\partial\theta}{\bf{e}}_{\theta}
+\frac{1}{r\sin\theta}\frac{\partial\psi}{\partial\varphi}{\bf{e}}_\varphi
\
\end{equation}
is the Laplacian in the spherical coordinates.  By adopting new
Cartesian coordinates $x,y,z$ within $\bf{e}_x\equiv
\bf{e}_\varphi$ and axis $z$ being local vertical, when the
stationary state is assumed to have the form
\begin{equation}
\psi(x)=\psi_\upsilon(z)e^{i(k_1x+k_2y)} \ ,
\end{equation}
one can easily derive from (\ref{Hamil}) that
\begin{eqnarray}
&& \hspace{-1cm}H_3\psi=i\hbar a c\omega
r\sin\theta\left(1-\sqrt{1+\frac{4GM}{\omega
c^2r^3}}\right)\frac{\partial\psi}{\partial x}=\nonumber\\
&& \hspace{-1cm} \hbar k_{1}ac\omega
r\sin\theta\left(1-\sqrt{1+\frac{4GM}{\omega c^2r^3}}\right)\psi
=-mu_1a c\omega r\sin\theta\left(1-\sqrt{1+\frac{4GM}{\omega
c^2r^3}}\right)\psi ,
\end{eqnarray}
where the following notation
\begin{equation}
u_1\equiv{\bf u}\cdot \mathbf{e}_\varphi, \quad {\bf
u}\equiv\frac{\hbar}{m}(k_1\mathbf{e}_x+k_2\mathbf{e}_y) \
\end{equation}
has been used.

 Following to the papers Ref.~\refcite{arminjon,vika} one can compute "KS parameter $\omega$"
modification of the energy level as the first-order perturbation:
\begin{equation}\label{var}
(\delta E)_{\omega}\simeq\langle\psi|H_3\psi\rangle=-mu_1\int
ac\omega r\sin\theta\left(1-\sqrt{1+\frac{4GM}{\omega
c^2r^3}}\right)|\psi|^2dV \ .
\end{equation}
Assume $r=(R+z)\cos\chi$ (where $\chi$ is the latitude angle) and
$\sin\theta$ to be equal to 1, that is $\theta=\pi/2$. Assuming
now $z\ll{ R}$ one can extend (\ref{var}) as
\begin{eqnarray} &&(\delta
E)_{\omega}\simeq -mu_1ac\omega
R\cos\chi\bigg[1-\sqrt{1+\frac{4GM}{c^2 \omega R^3\cos^3\chi}}
\nonumber\\ &&\qquad\quad
+\frac{6GM}{c^2 \omega R^4 cos^3\chi}\left(1+\frac{4GM}{c^2\omega
R^3\cos^3\chi}\right)^{-1/2}\int z|\psi|^2dV\bigg] \ ,
\end{eqnarray}
We remember that $\int z|\psi|^2dV$ is the average value $\langle
z\rangle_n$ of $z$ for the stationary state $\psi=\psi_n$. For the
further calculation we use formula for $\langle z\rangle_n$
from~\cite{arminjon}
\begin{equation}
\langle z\rangle_n=\frac{2}{3}\frac{E_n}{mg}\ .
\end{equation}
Now one can easily estimate the relative "KS parameter $\omega$"
modification of the energy level $E_n$ of the neutrons as
\begin{equation}
\frac{(\delta
E)_{\omega}}{E_n}\simeq-\frac{4u_1a}{cRcos^2\chi}\left(1+\frac{4g}{c^2\omega
R\cos^3\chi}\right)^{-1/2}
\end{equation}
We numerically estimate the obtained modification using the
typical parameters for the Earth: $u_1\simeq +10^3 \rm{cm/s}$,
$\omega\sim10^{-24} \rm{cm}^{2}$, $\cos\chi\simeq0.71$,
$a\simeq3.97\times 10^{2} \rm{cm}$,
 $g\simeq10^{3} \rm{cm/s}^2$ and $R\simeq6.4\cdot10^{8} \rm{cm}$, $c\simeq3\cdot10^{10} \rm{cm/s}$
%
\begin{equation}\label{numres}
\frac{(\delta E)_{\omega}}{E_n}\simeq2\times10^{-13}\ .
\end{equation}
%
 From the obtained result (\ref{numres}) one can see, that the in
influence of $\omega$ parameter will be stronger in the vicinity
of compact gravitating objects with small $R$. Recent experiments
on measuring energy levels of UCN~\cite{nesv} has an error $\sim
10^{-10}$, which does not allow to obtain the influence of
$\omega$ parameter on energy levels of UCN. Further improvements
of the experiments would give either exact value or lower limit
for the above mentioned parameter.

\section{The interference in a Mach-Zehnder-type
interferometer \label{3rdsec}}

The components of the tetrad frame for the proper observer for
metric (\ref{metrik}) are
\begin{eqnarray}
\label{zamo_tetrad_1} &&{\bf e}_{\hat
t}^{\mu}=
N^{-1}\left(1,0,0,0\right)\ ,  \quad {\bf e}_{\mu}^{\hat
t}=-
N\left(1,0,0,\frac{(1-N^2)a\sin^2\theta}{N^2}\right)\ , \nonumber
\\ \nonumber\\
\label{zamo tetrad 2} && {\bf e}_{\hat
r}^{\mu}=
N(0,1,0,0), \quad {\bf e}_{\mu}^{\hat
r}=
N^{-1}(0,1,0,0)\ . \\
\label{zamo tetrad 3} && {\bf e}_{\hat
\theta}^{\mu}=r^{-1}\left(0,0,1,0\right), \hskip 0.8cm {\bf
e}_{\mu}^{\hat \theta}=r(0,0,1,0)\ ,\nonumber \\
\nonumber\\
\label{zamo tetrad 4} && {\bf e}_{\hat
\varphi}^{\mu}=\frac{1}{r\sin\theta}\left(-\frac{(1-N^2)a\sin^{2}\theta}{N^2},0,0,1\right)\
, \quad  {\bf e}_{\mu}^{\hat \varphi}=r\sin\theta(0,0,0,1) \ ,
\end{eqnarray}
and the acceleration of the Killing trajectories~\cite{Valeria} is
\begin{eqnarray}
a_{\mu}=\frac{1}{2}\partial_{\mu}\ln(-g_{00}) \ ,
\end{eqnarray}
and we obtain for the nonvanishing component of the acceleration:
\begin{equation}
a_{\hat r}= \omega r\left(1-\frac{1+\frac{M}{\omega
r^3}}{\sqrt{1+\frac{4M}{\omega r^3}}}\right)\left(1+\omega
r^2-\omega r^2\sqrt{1+\frac{4M}{\omega r^3}}\right)^{-1/2}\ .
\end{equation}
The nonvanishing orthonormal ("hatted") components of rotation
tensor of the stationary congruence $\chi_{\mu \nu}$ in the slowly
rotating Ho\v{r}ava-Lifshitz gravity are given by

\begin{equation}
\chi_{\hat r \hat \varphi}=\frac{a\omega\sin\theta}{1+\omega
r^2-\omega r^2\sqrt{1+\frac{4M}{\omega
r^3}}}\left(1-\frac{1+\frac{M}{\omega
r^3}}{\sqrt{1+\frac{4M}{\omega r^3}}}\right) \ ,
\end{equation}
\begin{equation}
\chi_{\hat \theta \hat
\varphi}=\frac{a\omega\cos\theta}{\sqrt{1+\omega r^2-\omega
r^2\sqrt{1+\frac{4M}{\omega
r^3}}}}\left(1-\sqrt{1+\frac{4M}{\omega r^3}}\right)
 \ .
\end{equation}

The simple form of the vector potential of the electromagnetic
field $\mathcal{A_{\mu}}$  in the Lorentz gauge in the spacetime
(\ref{metrik}) is $\mathcal{A}^{\alpha}=
C_{1}\xi^{\alpha}_{t}+C_{2}\xi^{\alpha}_{\varphi}$~\cite{Ahmadjon}.
Here the integration constant $C_{2}=B/2$, where gravitational
source is immersed in the uniform magnetic field $\bf B$ being
parallel to its axis of rotation (properties of black hole
immersed in external magnetic field have been studied, for
example, in Ref.~\refcite{wald,aha
hl,Ahmadjon,ABprd,Konoplya:2006gg,Konoplya:2006qr}), and the other
integration constant $C_{1}=aB$ can be calculated from the
asymptotic properties of spacetime (\ref{metrik}) at the infinity:
\begin{eqnarray}
&&\mathcal{A}_{t}=-a B\left[1+\omega
r^2\left(1-\sqrt{1+\frac{4M}{\omega
r^3}}\right)\left(1-\frac{\sin^2\theta}{2}\right)\right] \ , \nonumber\\
&&\mathcal{A}_{\varphi}=\frac{Br^2}{2}\sin^2\theta \ .
\end{eqnarray}
One can write the total energy of the particle in the weak field
approximation in the following form:
\begin{eqnarray}
\mathcal{E}=p(\xi)+\mathcal{E}_{pot}=p(\xi)+e_{p}\mathcal{A}_{t} \
,
\end{eqnarray}
where $e_{p}$ is electric charge of the particle. This is
interpreted as total conserved energy consisting of
gravitationally modified kinetic and rest energy $p(\xi)$, a
modified electrostatic energy $e_{p}\mathcal{A}_{t}$.

  For the further use note the measured components of the
  electromagnetic field, which are the electric $E_{\alpha}=F_{\alpha\beta}u_{\rm{obs}}^{\beta}$
  and magnetic
  $B_{\alpha}=(1/2)\eta_{\alpha\beta\mu\nu}F^{\beta\mu}u_{\rm{obs}}^{\nu}$
  fields:
\begin{eqnarray}
\label{electormagnetic field 1} B_{\hat r}=-\frac{B\cos\theta}{2}
\ , \hskip 1cm B_{\hat
\theta}=-\frac{B\sin\theta}{2}\left(1+\omega r^2-\omega r^2\sqrt{1+\frac{4M}{\omega r^3}}\right)^{\frac{1}{2}} \ , \\
\label{electormagnetic field 2} E_{\hat r}=\frac{aB\omega
r}{\sqrt{1+\frac{4M}{\omega r^3}}}\left[2\left(1+\frac{M}{\omega
r^3}-\sqrt{1+\frac{4M}{\omega r^3}}\right)+\frac{3M}{\omega
r^3}\sin^2\theta\right], \hskip 0.4cm E_{\hat \theta}=0 \ ,
\end{eqnarray}
where
$F_{\alpha\beta}=\mathcal{A}_{\beta,\alpha}-\mathcal{A}_{\alpha,\beta}$
is the
  field tensor,
  $\eta_{\alpha\beta\mu\nu}=\sqrt{-g}\epsilon_{\alpha\beta\mu\nu}$
  is the pseudo-tensorial expression for the Levi-Civita symbol
  $\epsilon_{\alpha\beta\mu\nu}$, $g\equiv det|g_{\alpha\beta}|$.

Now one can obtain the total phase shift~\cite{Valeria} as
\begin{eqnarray}
&&\Delta\phi=\mathcal{E}S\bigg[-\frac{\mathcal{E}}{p_{0}}\left(\cos{\beta}a_{\hat
r } -\cos{\gamma}\sin{\beta}a_{\hat \theta} -\sin\gamma\sin\beta
a_{\hat \varphi }\right)
 \nonumber\\&&\qquad\qquad-\frac{1}{p_{0}}\left(\cos\beta\partial_{\hat
r}\mathcal{E}_{pot} -\cos{\gamma}\sin{\beta}\partial_{\hat
\theta}\mathcal{E}_{pot} -\sin{\gamma}\sin{\beta}\partial_{\hat
\varphi}\mathcal{E}_{pot}\right)
\nonumber\\&&\qquad\qquad+\sin\beta\chi_{\hat\theta \hat
\varphi}+\cos\gamma\cos\beta\chi_{\hat\varphi \hat
r}+\sin\gamma\cos\beta\chi_{\hat r \hat \theta} \bigg]\nonumber\\
 &&\qquad\qquad+e_{p}S\left(\sin\beta B_{\hat
r}+\cos\gamma\cos\beta B_{\hat \theta}+\sin\gamma\cos\beta B_{\hat
\varphi}\right) \ ,
\end{eqnarray}
where $\partial_{\hat \mu}=e_{\hat \mu}^{\nu}\partial_{\nu}$,
$\gamma$ is the angle of the baseline with respect to ${\bf
e}_{\hat \varphi}$ and $\beta$ is the tilt angle.
 Therefore one can independently vary the angles $\beta$ and
 $\gamma$, and extract from the phase shift measurements the
 following combinations of terms:
\begin{eqnarray}
&& \Delta\phi\left(\beta=0,\gamma=0\right)=\nonumber\\
&& \qquad
\mathcal{E}S\omega\mathcal{W}\left[\frac{a\sin\theta}{A^2}+\frac{a
B r e_{p}
A}{p_{0}}(2-\sin^2\theta)-\frac{\mathcal{E}r}{p_{0}A}\right]
-\frac{1}{2}e_pBS\sin\theta A, \qquad \\
&& \Delta\phi\left(\beta=\frac{\pi}{2},
\gamma=\frac{\pi}{2}\right)= S\cos\theta\left(\frac{a\omega
\mathcal{E}\mathcal{K}}{A}-\frac{e_{p}B}{2}\right), \\
&& \Delta\phi\left(\beta=\frac{\pi}{2},\gamma=0\right)=
S\cos\theta\left[\frac{a\omega\mathcal{E}
\mathcal{K}}{A}\left(1+\frac{Br\sin\theta
A}{p_{0}}\right)-\frac{e_{p}B}{2}\right], \\
&& \Delta\phi\left(\beta=0,\gamma=\frac{\pi}{2}\right)=-
\frac{\omega r S\mathcal{E}^2\mathcal{W}}{Ap_0}
\left[1-\frac{ae_{p}B A^2}{\mathcal{E}}(2-\sin^2\theta)\right],
\end{eqnarray}
where
\begin{eqnarray}
&&\left(1-\frac{1+\frac{M}{\omega r^3}}{\sqrt{1+\frac{4M}{\omega
r^3}}}\right)=\mathcal{W},\quad \left(1-\sqrt{1+\frac{4M}{\omega
r^3}}\right)=\mathcal{K}.\
\end{eqnarray}

Using above obtained results one can estimate lower limit for KS
parameter $\omega$. Using the results of the Earth based atom
interferometry experiments~\cite{kasevich} would give us an
estimate $\omega \simeq 1.25\times 10^{-25} \rm{cm}^{-2}$ .\

\section{The Sagnac effect in the Horava gravity \label{4thsec}}

\normalsize It is well known that the Sagnac effect for
counter-propagating beams of particles on a round trip in an
interferometer rotating in a flat space-time may be obtained by a
formal analogy with the Aharonov-Bohm effect. Here we study the
interference process of matter or light beams in the spacetime of
slowly rotating compact gravitating object in braneworld in terms
of the Aharonov-Bohm effect~\cite{ruggiero1}. The phase shift
\begin{eqnarray}\label{phaseshift}
&& \Delta\phi=\frac{2m u_0}{c\hbar}\oint_C {\bf A}_G\cdot d{\bf x}
\end{eqnarray}
is detected  at uniformly rotating interferometer and the time
difference between the propagation times of the co-rotating and
counter-rotating beams is equal to
\begin{eqnarray}\label{timedelay}
&& \Delta T=\frac{2u_t}{c^3}\oint_C {\bf A}_G\cdot d{\bf x} .
\end{eqnarray}

In the expressions (\ref{phaseshift}) -- (\ref{timedelay}) $m$
indicates the mass (or the energy) of the particle of the
interfering beams, ${\bf{A}}_G$ is the gravito-magnetic vector
potential which is obtained from the expression
\begin{eqnarray}\label{gravito}
({\bf A}^{G})_{i}\equiv c^2\frac{u_i}{u_0} ,
\end{eqnarray}
and $u^\alpha (x)$ is the unit four-velocity of particles:
\begin{equation}
u^\alpha\equiv\left\{\frac{1}{\sqrt{-g_{tt}}},0,0,0\right\}, \quad
u_\alpha\equiv\left\{-\sqrt{-g_{tt}},g_{it}/\sqrt{-g_{tt}}
\right\} \ .
\end{equation}

From (\ref{metrik}) and coordinate transformation
$\varphi\rightarrow\varphi+\Omega t$, where
$\Omega=\Omega_{\bigoplus}$ one can see that the unit vector field
$u^\alpha$ along the trajectories $r=R=\rm{const}$ will be
\begin{equation}
u_{t}=-(u^{t})^{-1}, \qquad u_{\phi}=[\Omega r^{2}-a(1-N^2)]u^{t},
\end{equation}
where we have used the following notation
\begin{equation}
u^{t}=\left[N^{2}-\Omega^{2}r^{2}+2a\Omega(1-N^2)\right]^{-1/2} \
.
\end{equation}
Now inserting the components of $u^{\alpha}$ into the equation
(\ref{gravito}) one can obtain
\begin{equation}
{\bf A}^{G}_{\phi}=-[\Omega r^{2}-a(1-N^2)](u^{t})^2.
\end{equation}

Integrating vector potential as it is shown in equations
(\ref{phaseshift}) and (\ref{timedelay}) one can get the following
expressions for $\Delta\phi$ and $\Delta T$ (here we returned to
the physical units):
\begin{equation}
\Delta\phi =\frac{4\pi
m}{\hbar}r^{2}(\Omega-\widetilde{\omega})u^{t}\ , \quad \Delta
T=\frac{4\pi}{c^{2}}r^{2}(\Omega-\widetilde{\omega}) u^{t}\ .
\label{ABDelta T}
\end{equation}
where $\widetilde{\omega}$ is the angular velocity of
Lense-Thirring.

Following to the paper Ref.~\refcite{ruggiero1} one can find a
critical angular velocity $\bar{\Omega}$
\begin{equation}
\bar{\Omega}=-a c\left(1-\sqrt{1+\frac{4M}{\omega r^3}}\right),
\end{equation}
which corresponds to zero time delay $\Delta T=0$. $\bar{\Omega}$
is the angular velocity of zero angular momentum observers (ZAMO).

\section{Conclusion}

We have studied quantum interference effects including e.g. the
phase shift and time delay in Sagnac effect  in the spacetime of
rotating gravitational objects in Ho\v{r}ava gravity and found
that they can be affected by the KS parameter $\omega$. Then, we
have derived an additional term  for the phase shift in a neutron
interferometer due to the presence of KS parameter and studied the
feasibility of its detection with the help of "figure-eight"
interferometer. We have also investigated the application of the
obtained results to the calculation of energy levels of UCN and
found modifications to be rather small for the Earth but more
relevant for the compact astrophysical objects. The result shows
that the phase shift for a Mach-–Zehnder interferometer in
spacetime of gravitational object in Ho\v{r}ava gravity is
influenced by the KS parameter $\omega$. Obtained results can be
further used in laboratory experiments to detect the interference
effects related to the phenomena of Ho\v{r}ava gravity. Recently
authors of the paper Ref.~\refcite{loboetal} from Solar system
tests obtained values for parameter $\omega$ as follow: $\omega
\simeq 3.1 \cdot 10^{-26} {\rm cm}^2$ (from perihelion precession
of the Mercury), $\omega \simeq 4.4 \cdot 10^{-26} {\rm cm}^2$
(light deflection by the Sun), $\omega \simeq 1.8 \cdot 10^{-25}
{\rm cm}^2$ (radar echo delay). Here we have estimated lower limit
for parameter $\omega$ as $\omega \simeq 2.4\times10^{-27}
\rm{cm}^{-2} \ $ using the experimental results of the recent
paper Ref.~\refcite{holgeretal} on the precise measurement of the
gravitational redshift by the interference of matter waves.

\section*{Acknowledgments}

The work was supported by the UzFFR (projects 5-08 and 29-08) and
projects FA-F2-F079 and FA-F2-F061 of the UzAS. This work is
partially supported by the ICTP through the OEA-PRJ-29 project.
Authors gratefully thank Viktoriya Morozova for useful
discussions. AB acknowledges the TWAS for associateship grant. AA
and AB thank the IUCAA for the hospitality where the research has
been completed.


\begin{thebibliography}{0}

%
\bibitem{h1}P. Ho\v{r}ava, \textit{JHEP} \textbf {0903},  020 (2009).
%

%
\bibitem{h2}P. Ho\v{r}ava, \textit{Phys. Rev. D} \textbf{ 79}, 084008  (2009).
%

%
\bibitem{Visser}
M. Visser, \textit{Phys. Rev D} \textbf {80}, 025011 (2009).
%

\bibitem{Sotiriou}
T. P. Sotiriou, M. Visser and S. Weinfurtner, \textit{Phys. Rev.
Lett.} \textbf{102}, 251601 (2009).
%
\bibitem{H_prl}
P. Ho\v{r}ava, \textit{Phys. Rev. Lett.} \textbf{102}, 161301
(2009).
%
\bibitem{Cai}
R. G. Cai, Y. Liu and Y. W. Sun, \textit{JHEP} \textbf{0906}, 010
(2009).
%
\bibitem{Orlando}
D. Orlando and S. Reffert, \textit{Class. Quant. Grav.}
\textbf{26}, 155021 (2009).
%
\bibitem{SotiriouJHEP}
T. P. Sotiriou, M. Visser and S. Weinfurtner, \textit{JHEP}
\textbf{0910}, 033 (2009).
%
\bibitem{Calcagni}
G. Calcagni, \textit{Phys. Rev. D}, \textbf{81}, 044006
(2010).
%
\bibitem{Germani}
C. Germani, A. Kehagias and K. Sfetsos, \textit{JHEP}, Issue 09,
060 (2009).

\bibitem{Takashi}

T. Takashi and J. Soda, \textit{Phys. Rev. Lett.} \textbf{102},
231301 (2009).
%
\bibitem{CalcagniJHEP}
G. Calcagni, \textit{JHEP} \textbf{09},112 (2009). ;
%
\bibitem{Kalyana}
S. Kalyana Rama, \textit{Phys. Rev. D} \textbf{79}, 124031 (2009).
%
\bibitem{Wang}
A. Wang and R. Maartens, \textit{Phys. Rev. D}, \textbf{81},
024009 (2010).

\bibitem{loboooo}{C. G. Boehmer, L. Hollenstein, F. S. Lobo and S. S. Seahra,
[arXiv:gr-qc/1001.1266].}
%
%
\bibitem{mukohayama}{S. Mukohyama, arXiv:1007.5199 (2010).}
%
\bibitem{Saridakis}
E. N. Saridakis, \textit{Eur. Phys. J. C} \textbf{67} 229 (2010).
%
\bibitem{parkkkk}
M. I. Park,     \textit{JCAP} \textbf{1001}, 001 (2010).
%
\bibitem{Mukohyama}
S. Mukohyama, \textit{Physical Review D} \textbf{80}, 064005
(2009).
%
%
\bibitem{loboetal}{T. Harko, Z Kovacs , F.S.N. Lobo, 2009arXiv0908.2874H . }
%
\bibitem{ghodsi}{A. Ghodsi, E. Hatefi, \textit{Physical Review D},
\textbf{81}, 044016 (2010). }
%
\bibitem{CaiCao}
R. G. Cai, L. M. Cao  and N. Ohta, \textit{Phys. Rev. D}
\textbf{80}, 024003 (2009).
%
\bibitem{Konoplya22}
R. A. Konoplya, \textit{Phys. Lett. B} \textbf{679}, 499 (2009).
%
\bibitem{SChen}
S. Chen and J. Jing, \textit{Phys. Rev. D} \textbf{80}, 024036
(2009).
%
\bibitem{CastilloLar}
A. Castillo and A. Larranaga, [arXiv:gr-qc/0906.4380].
%
\bibitem{ChenYang}
D. Y. Chen, H. Yang and X. T. Zu, \textit{Phys. Let B}
\textbf{681}, 463 (2009).
%
%
\bibitem{lobo1}
{F.S.N. Lobo, T. Harko and Z. Kova'cs, [arXiv:1001.3517v1
[gr-qc]}.
%
\bibitem{lobo2}
{C.G. B\"{o}hmer and F.S.N. Lobo, arXiv:0909.3986v2 [gr-qc]}.  
%
\bibitem{konoplya11}
{R. A. Konoplya, \textit{Phys. Lett. B} \textbf{679}, 499 (2009).
}
%
\bibitem{stuklik} Z. Stuchlik and A. Kotrlov\'{a}, \textit{Gen. Rel.
Grav.}, \textbf{41}, 1305 (2009).
%
\bibitem{stuklik2} A. Kotrlov\'{a}, Z. Stuchlik and G. T\"{o}r\"{o}k, \textit{Class.
Quantum Grav.} \textbf{25}, 225016 (2008).
%
\bibitem{aha hl} A.A. Abdujabbarov,  A.A.Hakimov and B.J. Ahmedov,
submitted (2010).
%
\bibitem{gwak}{B. Gwak and B.-H. Lee,     arXiv:1005.2805v2 (2010). }
%
\bibitem{overhauser}A.W. Overhauser and R.Colella, \textit{Phys. Rev.
Lett.}  \textbf{33}, 1237 (1974).
%
\bibitem{colella} R.Colella, A.W. Overhauser and S.A. Werner,
\textit{Phys. Rev. Lett.} \textbf{34}, 1472 (1975).

\bibitem{page} L.A. Page, \textit{Phys. Rev. Lett.} \textbf{35}, 543 (1975).

\bibitem{werner}S.A. Werner, J.L. Staudenmann and R.Colella,
\textit{Phys. Rev. Lett.} \textbf{42}, 1103 (1979).

\bibitem{mashhoon} B.Mashhoon, F.W. Hehl and D.S. Theiss, \textit{Gen.
Rel. Grav.} \textbf{16}, 711 (1984).

\bibitem{kuroiwa} J. Kuroiwa, M.Kasai and T. Futamase, \textit{Phys.
Lett. A} \textbf{182}, 330 (1993).

\bibitem{G.rizzi} G. Rizzi and M.L. Ruggiero,  gr-qc/0305084
(2004).

\bibitem{rizzi}G.Rizzi and M.L. Ruggiero, \textit{Gen. Rel. Grav.}
\textbf{35}, 1743 (2003).

\bibitem{ruggiero} M.L. Ruggiero, \textit{Gen. Rel. Grav.} \textbf{37},
1845 (2005).
%
\bibitem{nesvizhevsky} V. V. Nesvizhevsky et. al.,
\textit{Phys. Rev. D} \textbf{67}, 102002 (2003).
%
\bibitem{ks09} {A. Kehagias and K. Sfetsos, \textit{Phys. Lett. B} \textbf{678}, 123
(2009).}
%
\bibitem{four}{ D. Klemm, V. Moretti and L. Vanzo, \textit{Phys.
Rev. D} \textbf{57}, 6127 (1998). }
%

\bibitem{holgeretal}{H. Muller, A. Peters, S. Chu,
\textit{Nature} \textbf{463}, (2010) doi: 10.1038/nature08776. }
%
\bibitem{arminjon}M. Arminjon, \textit{Phys. Lett. A} \textbf{372}, 2196
(2008).

\bibitem{vika}V.S. Morozova and B.J. Ahmedov, \textit{Int. J. Mod. Phys.
D} \textbf{18}, 107 (2009).
%
\bibitem{nesv}{ C. Plonka-Spehr, A. Kraft, P. Iaydjiev, J. Klepp,
V.V. Nesvizhevsky, P. Geltenbort and Th. Lauer, \textit{Nucl.
Instr. M. Phys. R. A} \textbf{618}, 239 (2010). }
%
%
\bibitem{Valeria}V. Kagramanova, J. Kunz and C.
L\"{a}mmerzahl, \textit{Class. Quant. Grav.} \textbf{25}, 105023
(2008).
%
%
\bibitem{Ahmadjon}A. A. Abdujabbarov, B. J. Ahmedov and V. G.
Kagramanova, \textit{Gen. Rel. Grav.} \textbf{40}, 2515(2008).
%
\bibitem{ABprd}{A.A. Abdujabbarov, B.J. Ahmedov, \textit{Phys.
Rev. D}, \textbf{81}, 044022 (2010). }
%
\bibitem{Konoplya:2006gg}  R.~A.~Konoplya,  \textit{Phys. Lett. B} {\bf
644}, 219 (2007).
%
\bibitem{Konoplya:2006qr} R.~A.~Konoplya, \textit{Phys. Rev. D }{\bf
74}, 124015 (2006).
%
\bibitem{wald}{R.M. Wald, \textit{Phys. Rev. D}, \textbf{10}, 1680 (1974). }
%
\bibitem{kasevich} S. Dimopoulos, P.W. Graham, J.M. Hogan and M.E. Kasevich.
\textit{Phys. Rev. D}, \textbf{78}, 042003 (2008).

\bibitem{ruggiero1}M.L. Ruggiero, \textit{Gen. Rel. Grav.} \textbf{37},
1845 (2005).

\bibitem{bohmer} C.G. B\"{o}hmer, T. Harko and F. S. N. Lobo, \textit{Class. Quant.
Grav.}  {\bf 25}, 045015 (2008).

\bibitem{sepangi}S. Jalalzadeh., M. Mehrnia and H. R. Sepangi,
 \textit{Class. Quant. Grav.} \textbf{26}, 155007 (2009).
 %



\end{thebibliography}
\end{document}